# Video-rate volumetric chemical imaging via mid-infrared photothermal optical diffraction tomography

Masato Fukushima[1], Keiichiro Toda[2*], Yusei Sugawara[1], Shotaro Kawano[2], and Takuro Ideguchi[1,2,3**]

[1] Department of Physics, The University of Tokyo, Tokyo, Japan

[2] Institute for Photon Science and Technology, The University of Tokyo, Tokyo, Japan

[3] Optical Life Metrology Research Team, RIKEN Center for Advanced Photonics, RIKEN, Japan

* keiichiro--toda@g.ecc.u-tokyo.ac.jp

** ideguchi@ipst.s.u-tokyo.ac.jp

**Abstract**
Label-free vibrational microscopy provides chemically specific access to cellular structure, yet quantitative volumetric chemical dynamics in living cells remain largely inaccessible, particularly on subsecond timescales relevant to intracellular transport and structural reorganization. This limitation arises because most high-speed vibrational techniques rely on raster scanning, which constrains volumetric throughput to approximately one volume per second (vps). Although mid-infrared photothermal (MIP) imaging offers a pathway toward spatially parallel chemical detection, existing implementations have remained far below video-rate volumetric operation, reflecting a fundamental trade-off between imaging speed and signal-to-noise ratio. Here, we overcome this trade-off in MIP tomography and realize video-rate volumetric chemical imaging using mid-infrared photothermal optical diffraction tomography (MIP-ODT), achieving high photothermal sensitivity while maintaining quantitative measurement fidelity. High per-angle detectability supports volumetric reconstruction without temporal averaging, yielding a signal-to-noise ratio exceeding 70 under video-rate acquisition conditions. Consequently, volumetric imaging at 19.2 vps is achieved, representing a nearly 400-fold improvement over prior implementations. Using this capability, we performed video-rate three-dimensional tracking of lipid droplets in living cells and quantified anomalous diffusion from full volumetric trajectories, revealing heterogeneous intracellular transport behaviors that are obscured in two-dimensional measurements. We further demonstrate high-speed hyperspectral volumetric chemical imaging across a 300 cm$^{-1}$ spectral window within 1 s through rapid MIR wavenumber sweeping, paving the way for real-time three-dimensional organelle-specific chemical phenotyping. These results establish MIP-ODT as a platform for real-time, label-free volumetric chemical imaging and enable quantitative investigation of dynamic intracellular organization in heterogeneous cellular environments on biologically relevant timescales.

**Introduction**
In life science, volumetric imaging provides three-dimensional (3D) access to subcellular architecture and dynamic processes, supporting diverse biomedical studies spanning from organelle transport to mapping of tissue microenvironments[1,2]. Among existing live-cell imaging techniques, 3D fluorescence microscopy is widely regarded as the gold standard due to its high image contrast and molecular specificity[3]. However, fluorescence-based approaches rely on exogenous labels, which can perturb native biological states, complicate sample preparation, and restrict the number of molecular species that can be simultaneously interrogated. Label-free vibrational microscopy offers a complementary route by directly extracting chemically specific information from endogenous molecular vibrations. Coherent Raman microscopy, in particular, has been widely adopted because it enables rapid two-dimensional (2D) imaging with high spatial resolution, making it well suited for live-cell chemical imaging[4]. Extending coherent Raman microscopy to volumetric imaging has been achieved by incorporating axial scanning, enabling 3D chemical visualization in diverse contexts such as lipid droplet organization[5], myelin-rich neural tissue[6], and metabolic heterogeneity in multicellular assemblies[7]. Despite these advances, volumetric Raman imaging remains fundamentally constrained in throughput. The requirement for sufficient signal-to-noise ratio (SNR) typically necessitates point-scanning acquisition, imposing a fundamental trade-off between imaging speed and sampled volume. Consequently, even the fastest reported implementations are limited to whole-cell volumetric imaging rates of approximately one volume per second (vps), which is inadequate for resolving many intracellular dynamics that occur on sub-second timescales[8,9].

Mid-infrared photothermal (MIP) microscopy has recently emerged as a powerful approach for high-speed chemical imaging that overcomes the diffraction limit of mid-infrared (MIR) light[10-19]. In MIP microscopy, MIR absorption induces local refractive index (RI) changes that are detected by a visible probe beam, enabling bond-selective chemical imaging with visible diffraction-limited resolution. Because MIR absorption is a linear process, MIP microscopy can be implemented in both point-scanning and wide-field configurations, and recent advances have demonstrated video-rate 2D MIP imaging in both modalities[17,18]. In wide-field MIP, photothermal contrast is captured over the entire field of view within a single camera exposure. This spatial parallelism makes wide-field MIP, in principle, well suited for further speed scaling and suggests a promising pathway toward video-rate volumetric imaging. 3D information can be obtained by integrating wide-field MIP with diffraction tomography techniques, such as optical diffraction tomography (ODT)[14] or intensity diffraction tomography (IDT)[16], which reconstruct 3D RI distributions from amplitude-and-phase or intensity-only measurements acquired under multiple visible illumination angles. A key advantage of diffraction tomography is its flexible trade-off between imaging throughput and reconstruction fidelity. Imaging speed can be increased by reducing angular sampling density without decreasing the sampled volume, making this framework particularly attractive for high-speed volumetric chemical imaging. Despite this potential, previously reported implementations of wide-field MIP tomography have achieved volumetric imaging rate below 0.05 vps[14,16]. This limitation arises because existing approaches either suffer from insufficient single-frame SNR at each illumination angle, necessitating frame averaging, or are limited by the switching speed of the visible illumination angles. As a result, a substantial gap remains between current MIP-based tomographic methods and video-rate volumetric chemical imaging.

In this work, we establish a fundamentally distinct performance regime in MIP tomography by jointly advancing per-frame MIP detection sensitivity and angular switching speed, thereby transcending the conventional speed-sensitivity trade-off and enabling video-rate volumetric chemical imaging. We implement this strategy within a MIP-ODT framework. High MIP sensitivity is achieved by employing a high-power nanosecond-pulsed MIR source generated by an optical parametric oscillator (OPO), which provides sufficient SNR within a single camera exposure at each illumination angle. Rapid angular scanning is realized through fast wavefront modulation using a high-speed spatial light modulator (SLM). Using eleven illumination angles with a numerical aperture (NA) of 0.85, our system achieves a volumetric imaging rate of 19.2 vps, enabling video-rate 3D tracking of lipid droplets in living cells and quantitative analysis of their anomalous diffusion. Furthermore, integration of a fast MIR wavenumber-swept module enables 3D hyperspectral imaging, allowing the acquisition of 20 spectral points spanning a 300 $cm^{-1}$ window within approximately one second. Using this capability, we resolved distinct vibrational spectral signatures associated with subcellular structures, including nucleoli and lipid droplets. Together, these advances establish a platform for real-time, label-free volumetric chemical imaging at the single-cell level and open new opportunities for visualizing dynamic intracellular molecular processes, including drug accumulation, lipid droplet formation and fusion, and the emergence of intracellular aggregates.

## Results
### Video-rate MIP-ODT system.
We first outline the principle of MIP-ODT. In this method, wide-field MIR irradiation induces localized heating across the field of view through MIR absorption, generating transient RI changes via the thermo-optic effect. These induced RI changes are quantitatively measured by ODT, where 2D optical wavefronts are interferometrically captured under multiple visible illumination angles with and without MIR excitation (Fig. 1a). To reconstruct 3D distribution of the MIP-induced RI change from the measured wavefront images, the wavefront modulation is extracted for each illumination angle from the difference between MIR-ON and MIR-OFF measurements. Under the weak-scattering approximation, this differential wavefront is mapped to the corresponding spatial-frequency components of the RI change in the sample's 3D Fourier space. By assembling the spatial-frequency information across all illumination angles, we reconstruct the 3D distribution of MIP-induced RI changes (see "Procedure of RI reconstruction" in Methods for details).

We then designed a MIP-ODT system for video-rate volumetric imaging. To determine an appropriate balance between imaging speed and reconstruction fidelity, we evaluated reconstructed RI images obtained with different numbers of visible illumination angles (Fig. 1b, see "ODT measurement with a variable number of illuminations" in Methods). Based on this analysis, the number of illumination angles was set to eleven to maximize volumetric acquisition speed while maintaining sufficient image quality to capture global cellular morphology. Figure 1c provides a timing diagram of the video-rate MIP-ODT system. High-speed angular scanning of visible illumination is achieved through fast wavefront control using a high-speed SLM operating at 422.4 Hz, corresponding to an angular update period of 2.3 ms. Under these conditions, the volumetric MIP acquisition rate is given by $v_{ps} = 422.4/(2 \cdot 11)$=19.2 vps, which corresponds to an effective volumetric acquisition time of 52 ms. Here, the factor of two accounts for the alternating acquisition of MIR-ON and MIR-OFF frames. The system supports two distinct operational modes (Fig. 1d). In the video-rate single-wavenumber mode, volumetric imaging is performed at the full video-rate acquisition speed. In the high-speed hyperspectral mode, rapid MIR wavenumber sweeping enables the acquisition of 19.2 spectral points per second. Given that the intrinsic spectral linewidth of the MIR excitation is approximately 10 cm$^{-1}$ [18], this configuration enables scanning of a spectral window of several hundred cm$^{-1}$ within one second.

Figure 1e shows the optical layout of the video-rate MIP-ODT system. The MIR and visible pulses are derived from two electronically synchronized Q-switched Nd:YAG lasers (wavelength 1,064 nm; pulse width, 10 ns; repetition rate, 422.4 Hz; NL202, Ekspla). The visible probe beam is generated via second-harmonic generation (SHG) using a 15-mm-long LBO crystal. The MIR pump beam is provided by the idler output of an OPO incorporating a fan-out periodically poled lithium niobate (PPLN) crystal as the nonlinear medium (HC Photonics Corporation). The MIR output wavenumber is continuously tunable over the range 2,400-3,600 cm$^{-1}$ by laterally translating the fan-out PPLN crystal using a motorized translation stage (LNR25ZFS/M, Thorlabs), which selectively accesses different poling periods along the crystal. The OPO delivers pulse energies of ~3 µJ at the sample plane, more than one order of magnitude higher than those of quantum cascade laser (QCL) sources used in previous MIP tomography implementations[14,16]. This substantially enhances the single-frame MIP detection sensitivity. The short 10-ns MIR pulses effectively suppress spatial blurring caused by thermal diffusion, limiting heat-induced resolution degradation to ~10 nm[19]. In addition, the sub-kHz repetition rate ensures sufficient thermal relaxation between successive excitation pulses, thereby minimizing thermal pile-up. This reduces the risk of heat-induced damage during prolonged live-cell imaging and preserves the quantitative accuracy of the measured RI change signals.

The visible probe beam is directed into a single-objective ODT system based on off-axis digital holography (DH) implemented in a Mach-Zehnder interferometer[19]. In the sample arm, the illumination angle of the visible probe is rapidly modulated using a high-speed SLM (1920 × 1152 XY Phase Series Spatial Light Modulator, Meadowlark Optics) (see "Tilted illumination using an SLM" in Methods). After transmission through the specimen, the probe beam is reflected by a silicon substrate and recollected by the same objective lens (UPlanSApo60x, Olympus), forming a reflective configuration. This geometry allows MIR excitation to be introduced from the opposite side through the substrate. The NAs for visible illumination and detection are 0.85 and 1.2, respectively. The reflected visible beam is subsequently interfered with the reference beam at an off-axis angle, and the resulting holograms are recorded at 422.4 fps using a high-speed image sensor (VLXT-17M.I, Baumer). Camera acquisition is synchronized with the SLM modulation to ensure precise angular sampling. Details of the timing synchronization scheme are provided in Supplementary Note 1.

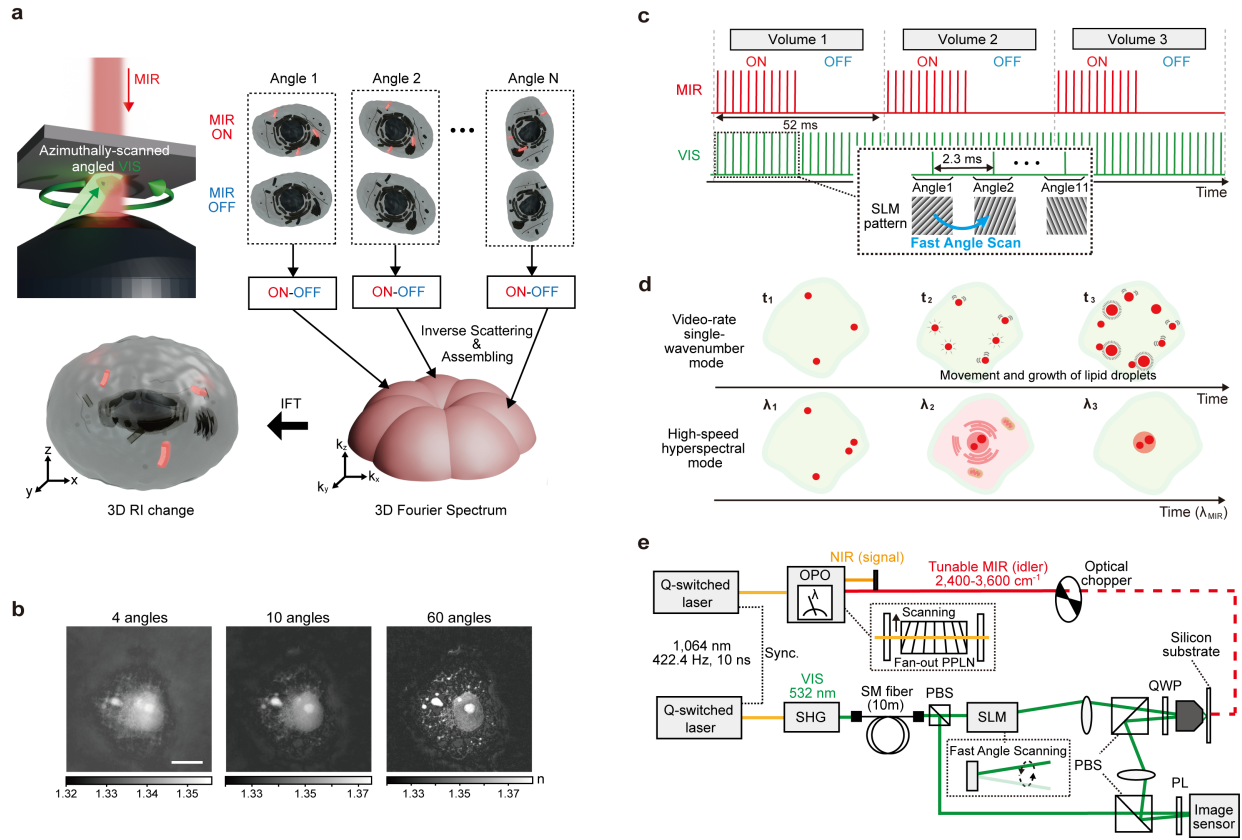

**Fig. 1 | Video-rate MIP-ODT system. a** Schematic of the MIP-ODT principle. IFT, Inverse Fourier transformation. **b** Reconstructed RI images obtained using different numbers of illumination angles. Scale bar, 10 μm. **c** Timing diagram of video-rate volumetric image acquisition. **d** Operating modes of the system: video-rate single-wavenumber mode and high-speed hyperspectral mode. **e** Optical layout of the video-rate MIP-ODT system. The MIR beam is modulated on and off using an optical chopper. VIS, visible; NIR, near-infrared; SM, single-mode; PBS, polarizing beam splitter; QWP, quarter-wave plate; PL, polarizer.

**Validation of the spatial resolution and signal-to-noise ratio (SNR).**

To validate the spatial resolution and SNR of video-rate MIP-ODT, we performed volumetric imaging of a living COS-7 cell (see "Preparation of biological samples" in Methods). Figure 2a shows representative depth-resolved slices of the RI tomograms (top) and the corresponding MIP images (bottom), acquired at a volumetric imaging rate of 19.2 vps. For MIP-ODT, a MIR pump wavenumber of 2,920 cm$^{-1}$ was used to excite the $CH_2$ asymmetric stretching vibration, which is abundant in lipid-rich cellular structures. In both ODT and MIP-ODT, depth-resolved slices reveal plane-specific 3D cellular features that vary with axial position, confirming successful volumetric reconstruction under video-rate acquisition conditions.

We first quantified the measurement noise to determine the SNR. The temporal RI fluctuations were measured from ODT data acquired in the MIR-OFF state. Averaging the voxel-wise standard deviations across the field of view yielded an effective noise level of $4.2 \times 10^{-5}$ (Supplementary Note 2). Using a representative peak MIP-induced RI change of $3.0 \times 10^{-3}$ from Fig. 2a as the signal amplitude, we obtain a signal-to-noise ratio of 71.4. This value indicates that a high signal-to-noise ratio is maintained under video-rate volumetric acquisition.

Next, spatial resolution was evaluated by analyzing the full width at half maximum (FWHM) of cross-sectional profiles extracted from a small intracellular particle, as indicated by white arrows in Fig. 2a, in both the reconstructed RI and the

MIP-induced RI change. Cross-sectional profiles along the lateral (x and y) and axial (z) directions are presented in Fig. 2b. Each profile was fitted with a Gaussian function to extract the corresponding FWHM values. For ODT, the measured FWHMs were 358 nm, 339 nm, and 957 nm along the x, y, and z axes, respectively. For MIP-ODT, the corresponding values were 361 nm, 337 nm, and 1.15 µm. These values are taken as representative lateral and axial spatial resolutions of the present system. Numerical simulations predict theoretical FWHM of 230 nm laterally and 870 nm axially for the point spread function. The slightly larger experimental values likely arise from the finite size of the intracellular particle used for evaluation, suggesting that the intrinsic resolution of the optical system may be modestly better than the measured estimates.

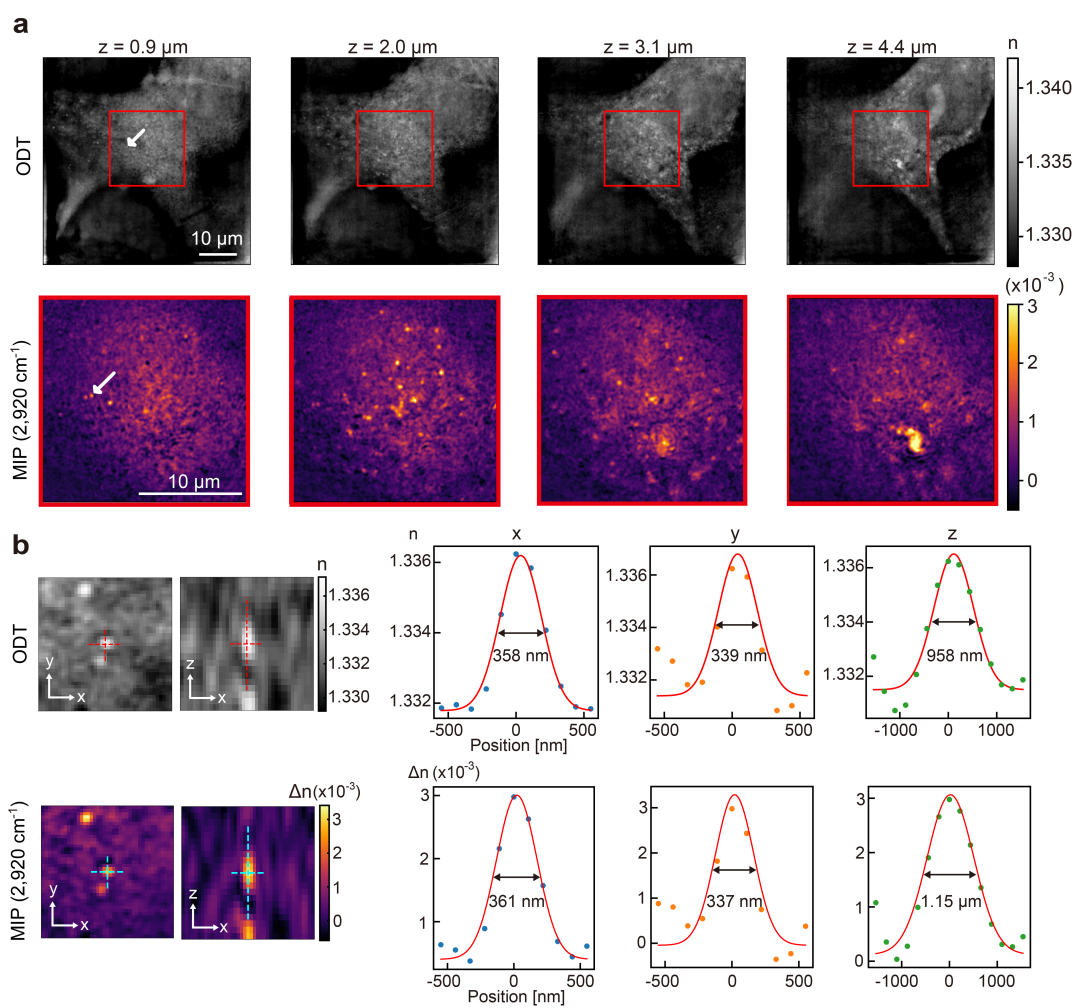

**Fig. 2 | Video-rate volumetric MIP-ODT imaging and spatial resolution characterization in a living cell. a** Depth-resolved slices extracted from the reconstructed RI tomogram of a living COS-7 cell (top). Corresponding MIP-induced RI change images acquired from the red-boxed regions, obtained at a volumetric imaging rate of 19.2 vps using a MIR pump tuned to 2,920 cm$^{-1}$ (bottom). White arrows indicate a small intracellular particle selected for spatial resolution analysis. The z-axis is aligned with the optical axis, and the origin is defined as the surface of the silicon substrate. The FOV of ODT and MIP-ODT are 55 × 55 µm$^2$ and 20 × 20 µm$^2$, respectively. The sample extent of MIP-ODT was estimated to be 10.1 µm along the axial direction. **b** Left, zoom-in views of the selected particle in the reconstructed RI (top) and MIP-induced RI change (bottom). Right, corresponding lateral (x, y) and axial (z) cross-sectional profiles extracted along the dashed lines in the left panels. Red solid lines represent Gaussian fits used to determine the FWHM.

**Real-time hyperspectral volumetric imaging.**

We next demonstrate real-time hyperspectral volumetric imaging by integrating the video-rate MIP-ODT platform with rapid wavenumber sweeping of the MIR excitation. This implementation enables the acquisition of up to 19.2 spectral sampling points per second within a selected spectral window. In this experiment, we focus on the 2,800-3,100 cm$^{-1}$ region, which includes the $CH_2$ and $CH_3$ stretching vibrations that are characteristic of lipids and proteins. At each volume, the instantaneous MIR excitation wavenumber, corresponding to the idler output of the OPO, is determined by measuring the wavelength of the counterpart near-infrared (NIR) signal beam using a spectrometer (C11482GA, Hamamatsu). This approach allows volume-by-volume spectral assignment under fully synchronized system control (see "Hyperspectral volumetric analysis" in Methods). Supplementary Note 3 summarizes the MIR wavenumber trajectories inferred from the measured NIR spectra for both unidirectional fast-sweeping and continuous-sweeping modes. These measurements confirm that spectral windows of approximately 300 cm$^{-1}$ and 200 cm$^{-1}$ can be covered within one second under unidirectional and continuous sweeping modes, respectively.

To validate 3D, location-dependent spectral readout in cells, we performed hyperspectral volumetric imaging of a fixed COS-7 cell immersed in $D_2O$-based phosphate-buffered saline (PBS). Figure 3a and 3b present representative depth-resolved slices of the RI tomograms and the corresponding MIP images, respectively. The MIP images shown in Fig. 3b were acquired at excitation wavenumbers of 2,932 cm$^{-1}$ and 2,962 cm$^{-1}$, which primarily probe $CH_2$- and $CH_3$-stretching vibrations and therefore provide chemical contrast associated with lipid-enriched and protein-enriched regions. Based on the volumetric reconstructions, we selected three representative subcellular regions of interest: a small cytoplasmic particle, a perinuclear structure, and the nucleolus. Local vibrational spectra were extracted from each region, as shown in Fig. 3c. The spectrum of the small particle exhibits pronounced $CH_2$-associated peaks around 2,860 cm$^{-1}$ and 2,920 cm$^{-1}$, consistent with lipid-rich compartments and supporting its assignment as a lipid droplet. In contrast, the perinuclear structure shows a relatively enhanced $CH_3$ contribution at 2,962 cm$^{-1}$ compared with the small particle, suggesting a higher protein fraction superimposed on lipid content. The spectrum extracted from nucleolus is dominated by a peak centered near 2,962 cm$^{-1}$, consistent with its well-known protein-rich molecular composition[14]. Together, these results demonstrate that hyperspectral MIP-ODT resolves distinct, compartment-specific vibrational signatures in 3D. By combining volumetric imaging with rapid spectral acquisition, this approach enables chemical phenotyping of intracellular architecture on a timescale of seconds.

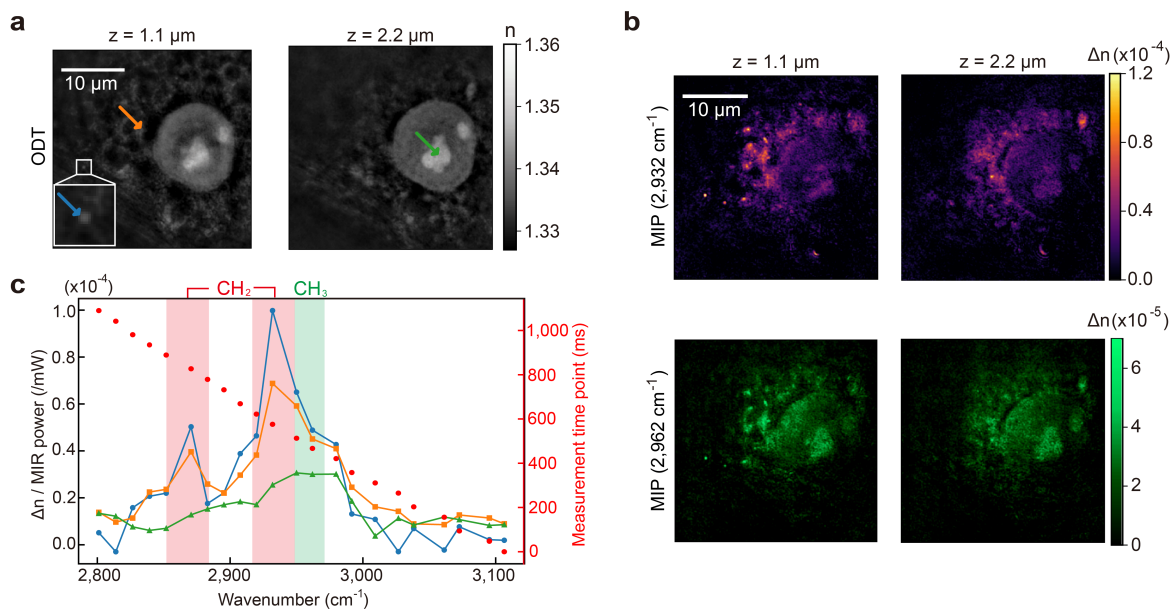

**Fig. 3 | Real-time hyperspectral volumetric MIP-ODT imaging of subcellular chemical heterogeneity. a** Representative depth-resolved slices from RI tomograms of a fixed COS-7 cell at two axial positions. **b** Corresponding MIP images acquired at 2,932 cm$^{-1}$ (top) and 2,962 cm$^{-1}$ (bottom). The FOV of ODT and MIP-ODT were 44 × 44 µm$^2$. The sample extent of MIP-ODT was estimated to be 2.4 µm along the axial direction. **c** Local vibrational spectra extracted

from three representative subcellular regions marked by color-coded arrows in **a**: a small cytoplasmic particle (blue), a perinuclear structure (orange), and the nucleolus (green). Shaded bands indicate the $CH_2$- and $CH_3$-stretching regions. Red dots represent the instantaneous MIR excitation wavenumber at each measurement time point during the wavenumber sweep.

**Video-rate 3D tracking of lipid droplets in living cells.**
To demonstrate video-rate 3D tracking of lipid droplets in living cells, we performed volumetric MIP-ODT imaging of living COS-7 cells treated with oleic acid, a condition known to robustly induce lipid droplet formation[20]. Volumetric datasets were acquired continuously for approximately 3.6 s at 19.2 vps, with the MIR excitation tuned to 2,920 cm$^{-1}$ to selectively probe lipid-rich structures. Representative depth-resolved slices are shown in Fig. 4a. Using this dataset, we analyzed the motion of an individual lipid droplet exhibiting pronounced displacement during the acquisition period, as indicated by the orange arrow (see "Motion analysis in video-rate MIP-ODT imaging" in Methods for details). The centroid position was tracked over time to reconstruct its full 3D trajectory (Fig. 4b). The tracking results overlaid on the MIP-ODT data are shown in Supplementary Video 1. The displacement spans approximately 200 to 300 nm along each of the x, y, and z axes. Such axial motion cannot be captured by conventional 2D imaging and would therefore lead to systematic underestimation of intracellular transport dynamics. In contrast, our volumetric imaging enables direct quantification of complete 3D motion within the cellular environment. 3D tracking of lipid droplets has previously been reported using ODT[21], in which droplets were identified from high-RI puncta. However, this contrast mechanism lacks intrinsic chemical specificity and may therefore require orthogonal validation. By contrast, MIP-ODT directly tracks lipid-rich structures via vibrational contrast, enabling chemically specific and volumetrically resolved characterization of droplet dynamics.

To quantitatively characterize the observed motion, we analyzed the centroid trajectory within the framework of anomalous diffusion. The mean squared displacement (MSD), calculated from the 3D trajectory over the 3.6 s observation window and evaluated for time lags $\tau$ up to 1.1 s, exhibits a power-law dependence on $\tau$ (Fig. 4c). Fitting the MSD yields an anomalous diffusion exponent of $\alpha = 0.32$ (95% bootstrap confidence interval (CI): 0.23–0.42, 500 bootstrap resamples), which is substantially lower than the value expected for free diffusion ($\alpha = 1$) and indicates pronounced subdiffusive behavior. Such subdiffusion is commonly observed for endogenous cargos in living cells and is generally attributed to molecular crowding and viscoelastic constraints that impede long-range transport[22].

We next extended this analysis to multiple lipid droplets within the same field of view. Figure 4d summarizes the distributions of the generalized diffusion coefficient $D_\alpha$ and the diffusion exponent $\alpha$ estimated from 50 tracked particles. The diffusion coefficients span approximately $10^{-3}$ to $10^{-2}$ $\mu m^2$ s$^{-\alpha}$, revealing marked heterogeneity in intracellular lipid droplet mobility. A previous study has reported anomalous diffusion exponents for intracellular lipid droplets spanning a broad range from approximately 0 to 2[21]. In contrast, the $\alpha$ values observed here are confined to a narrower range of 0.0 to 0.4. The particle-wise estimation uncertainty was moderate (median 95% bootstrap CI width for $\alpha$: 0.17, median relative 95% bootstrap CI width for $D_\alpha$: 0.17, 500 bootstrap resamples). This restricted distribution may reflect the relatively low metabolic activity of the cells under our experimental conditions, which would suppress ATP-dependent active processes, thereby biasing subcellular motion toward more locally constrained dynamics.

Finally, to probe the spatial organization of intracellular mobility, we mapped the spatial distribution of diffusion exponents $\alpha$ across the field of view (Fig. 4e). A subset of lipid droplets marked by green and orange arrows in Fig. 4a and 4e exhibited locally elevated $\alpha$ values, revealing pronounced spatial heterogeneity in diffusion behavior within the cell. These particles are located distal to the nucleus (dashed red circle in Fig. 4a), where intracellular crowding is reduced relative to perinuclear regions enriched in endoplasmic reticulum and other organelles. These observations suggest that lipid droplets experience less constrained motion in these areas. Collectively, these results demonstrate that video-rate MIP-ODT enables real-time, 3D tracking of chemically resolved intracellular structures, providing quantitative access to heterogeneous and potentially anisotropic particle dynamics in living cells.

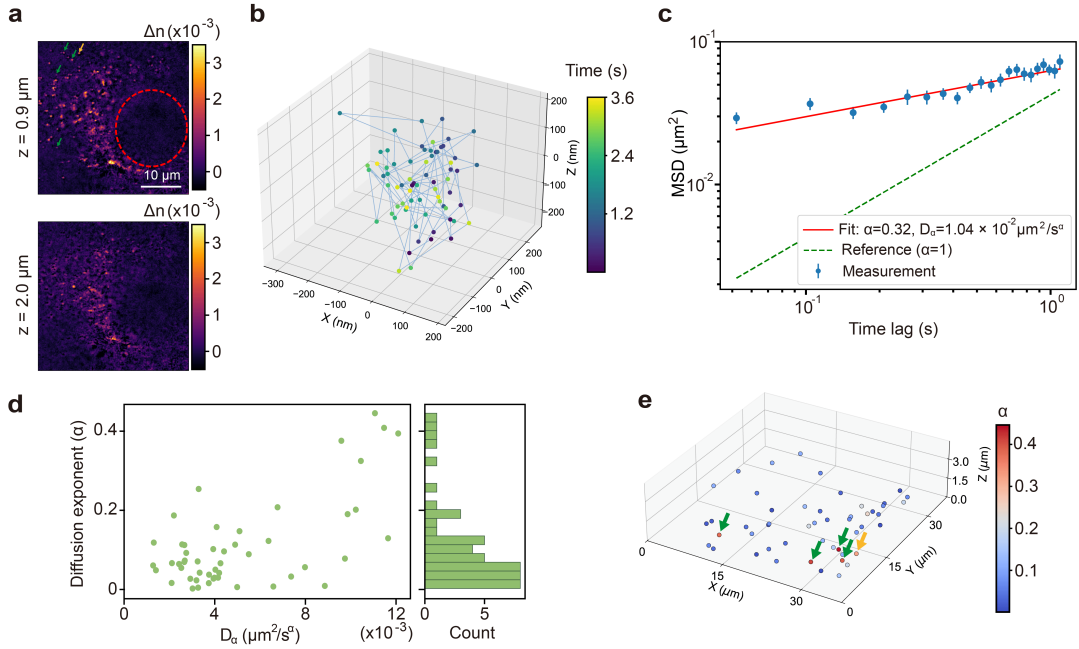

**Fig. 4 | Video-rate 3D tracking and anomalous diffusion of lipid droplets in living cells. a** Representative depth-resolved MIP-ODT slices of an oleic-acid-treated COS-7 cell acquired at 19.2 volumes per second with MIR excitation tuned to 2,920 cm$^{-1}$. The FOV of MIP-ODT is 37.4 × 37.4 µm$^2$. The sample extent of MIP-ODT was estimated to be 6.4 µm along the axial direction. **b** Reconstructed 3D trajectory of a lipid droplet indicated by the orange arrow in **a**, tracked over a 3.6 s acquisition window. **c** MSD calculated from the 3D trajectory as a function of time lag τ. **d** Scatter plot of the generalized diffusion coefficient $D_\alpha$ versus the anomalous diffusion exponent $\alpha$ obtained from MSD analysis of individual lipid-droplet trajectories in the field of view. The histogram on the right shows the distribution of $\alpha$ across trajectories (counts per bin). **e** Spatial map of anomalous diffusion exponents $\alpha$. Each color (green/orange) arrow corresponds to the same color arrow in **a**.

## Discussion

There are several directions for further technical development of MIP-ODT. A central opportunity is extending SLM-based wavefront control toward fully programmable illumination. In the current implementation, video-rate volumetric imaging employs a fixed set of eleven illumination angles. Programmable illumination would enable adaptive control of angular sampling density according to the characteristic timescale of the dynamics under investigation. This flexibility provides a tunable trade-off between reconstruction fidelity and acquisition speed, as illustrated in Fig. 1b. For example, when observing biological processes evolving on second timescales, the number of illumination angles can be increased by approximately an order of magnitude, thereby achieving reconstruction fidelity comparable to that obtained with 60 illumination angles in Fig. 1b, provided that strict video-rate acquisition is not required.

Programmable illumination also enables more advanced acquisition strategies. The SLM can generate complex illumination patterns that implement angular multiplexing within a single camera exposure, in combination with an appropriate forward model and model-based inversion[24]. Encoding multiple illumination angles into a single measurement expands Fourier-space coverage per exposure, thereby improving the fidelity of reconstructed MIP-induced RI-changes while preserving the volumetric imaging rate. Alternatively, angular multiplexing can maintain the same equivalent angular density while reducing the total number of measurement frames. This strategy enables several-fold increases in volumetric acquisition speed without compromising reconstruction fidelity.

We further consider resolution enhancement enabled by the reflective ODT geometry adopted here. Compared with conventional transmission-based implementations, this geometry provides access to higher illumination NAs[19]. Increasing the illumination and collection NAs to 1.2 and 1.3 is expected to improve spatial resolution to 168 nm laterally and 242 nm axially. Additional improvement can be achieved by shortening the probe wavelength. A 400 nm probe would yield estimated resolutions of 126 nm laterally and 181 nm axially, enabling 3D resolution of sub-200-nm structures. This resolution regime would permit label-free visualization of intracellular nanostructures that remain challenging for conventional 3D chemical imaging, including small vesicular compartments and membrane-associated nanostructures.

Finally, we discuss potential applications. This platform enables 3D imaging of intracellular dynamics on sub-second to second timescales with chemical specificity. Such capability enables direct investigation of early stages of lipid droplet biogenesis, during which nascent droplets emerge from the endoplasmic reticulum and grow following fatty-acid loading, as well as subsequent processes such as droplet growth and fusion in living cells[25]. More broadly, high-speed 3D chemical imaging provides a powerful approach to probe the formation and remodeling of biomolecular condensates driven by liquid-liquid phase separation, including nucleolar reorganization and stress-induced assemblies whose molecular components exchange on timescales of seconds to minutes[26]. Collectively, these perspectives position MIP-ODT as a versatile platform for real-time, label-free volumetric chemical imaging and enable quantitative investigation of dynamic intracellular organization and transport.

## Methods
### Procedure of RI reconstruction.
3D RI reconstruction was performed using the standard ODT workflow under the Rytov approximation[27]. The MIP-ODT reconstruction workflow was modified relative to conventional approaches. In conventional implementations, RI distributions are independently reconstructed for the MIR-ON and MIR-OFF states from complex optical fields referenced to background images measured in cell-free regions. The MIP-induced RI change is then obtained by subtracting the reconstructed MIR-OFF distribution from the MIR-ON distribution.

In this study, we introduced a direct differential reconstruction method that directly estimates the MIP-induced RI change from the differential complex optical fields between the MIR-ON and MIR-OFF states. This strategy removes the need for separate background acquisition in cell-free regions, thereby reducing experimental overhead and supporting real-time MIP imaging. A detailed formulation of the reconstruction method is provided in Supplementary Note 4.

The reconstructed ODT and MIP-ODT tomograms span (X) 58.7 × (Y) 58.7 × (Z) 29.3 μm$^3$. To focus on the MIR-irradiated region, we analyzed cropped subvolumes for each experiment. The cropped volumes are reported in the corresponding figure captions. The sample extent was estimated as the apparent axial range over which the lipid droplet contrast was observable, measured from the silicon substrate surface.

### ODT measurement with a variable number of illuminations.
To evaluate the effect of angular sampling density on reconstruction performance, we performed ODT measurements using a variable number of illumination angles, shown in Fig. 1b. Although this dataset was acquired using an optical configuration that differs from that in Fig. 1e, these modifications do not affect the conclusion of this analysis. In this configuration, a 1 ns visible pulse source was employed instead of the 10 ns laser used in Fig. 1e. In addition, the SLM was replaced with a wedge prism mounted on a rotation stage. The illumination NA was fixed at 1.0. Controlled azimuthal rotation of the prism enabled precise tuning of the illumination angle while maintaining a constant illumination NA.

### Tilted illumination using an SLM.
Tilted illumination was generated using an SLM programmed with a blazed phase grating pattern. This grating efficiently directs the incident optical power into the first diffraction order, which was used as the tilted illumination beam. To suppress undesired diffraction components, including the zeroth order and higher orders, spatial filtering was applied using an aperture mask placed at a Fourier plane. As a result, only the desired coherent tilted illumination was delivered to the specimen, enabling controlled angular illumination for volumetric imaging (Supplementary Note 5).

Because the SLM is operated with an alternating-current (AC) voltage, small periodic phase fluctuations are inherently introduced into the modulated wavefront. To mitigate the resulting phase noise, the timing of the MIR pump pulses, the visible probe pulses, and the camera exposure was synchronized to the AC driving cycle of the SLM.

**Preparation of biological samples.**
COS-7 cells are cultured on a silicon substrate with a thickness of 675 μm (6-675P1UMM, AKD, Japan). Cells were maintained in high-glucose Dulbecco's modified Eagle medium (DMEM; FUJIFILM Wako) containing L-glutamine, phenol red, and HEPES, supplemented with 10% fetal bovine serum (FBS; Cosmo Bio) and 1% penicillin-streptomycin-L-glutamine solution (FUJIFILM Wako). Cell cultures were maintained at 37 °C in a humidified incubator with 5% $CO_2$.

Prior to imaging, samples were mounted by placing a glass coverslip directly on top of the cells cultured on the silicon substrate. For the experiment shown in Fig. 3, cells were chemically fixed with 4% paraformaldehyde, after which the culture medium was replaced with $D_2O$-based PBS for imaging. For the live-cell experiment shown in Fig. 4, lipid-enriched cells were prepared by supplementing the culture medium with 100-200 μM of oleic acid and bovine serum albumin complex, followed by overnight incubation.

**Hyperspectral volumetric analysis.**
Hyperspectral MIP-ODT datasets were acquired by synchronizing volumetric image acquisition with the spectrometer readout using an index-based pairing scheme. A detailed description of the timing chart is provided in Fig. S5a in Supplementary Note 6. At the beginning of the measurement, the MIR beam was blocked using a mechanical shutter, and baseline MIR-OFF signals were recorded by both the camera and the spectrometer. After a short delay, the shutter was opened, producing a step-like increase in the MIR-induced signal. This shutter-opening time point was identified during post-processing and defined as the temporal origin of the hyperspectral measurement. Volumetric datasets acquired after the temporal origin were paired with the corresponding spectrometer records by matching their sequential acquisition indices.

The MIR idler wavenumber associated with each volume was calculated from the measured NIR signal wavenumber using the energy conservation relation of the OPO, $\omega_{pump} = \omega_{signal} + \omega_{idler}$. To correct for wavenumber-dependent variations in MIR excitation power during hyperspectral reconstruction, each MIP-ODT volume was normalized by the measured MIR pulse energy at the corresponding wavenumber. Because the MIR wavenumber is scanned continuously during each volumetric acquisition, the MIR pulse energy can vary slightly even within a single volume. Over the scanned spectral range, however, the MIR power dependence on wavenumber is smooth, and the resulting intra-volume variation is limited to at most 4.6% in relative intensity (Fig. S5b). We therefore expect spectral distortion introduced by this normalization to be negligible.

**Motion analysis in video-rate MIP-ODT imaging.**
Lipid droplets were detected and tracked using a supervised image-analysis pipeline. Lipid-droplet candidates were identified by the Pixel Classification in *ilastik* library[28]. The classifier was trained on manually curated in-house annotations that included only well-isolated droplets. Because reliable MSD analysis requires non-overlapping single-particle trajectories, droplets located in densely packed or clustered regions were deliberately excluded from the training set. As a result, the detection pipeline preferentially identifies spatially separated droplets and may under-represent droplets in high-density regions. For each reconstructed 3D volume, the lipid-droplet probability channel was extracted from the *ilastik* output and thresholded at $p \geq 0.8$. The thresholded volumes were subjected to 3D connected-component labeling. Components smaller than 5 voxels were discarded. Particle positions were defined as probability-weighted centroids of the remaining components. Trajectory reconstruction was performed using the *trackpy* library[29]. Tracking parameters were empirically optimized to suppress false detections and incorrect linking events. The maximum allowable inter-frame displacement was limited to 0.5 μm. Trajectories shorter than 60 frames out of a total of 70 frames were excluded to ensure robust MSD estimation.

For each 3D trajectory, MSD was calculated as $MSD(\tau) = \langle |r(t+\tau) - r(t)|^2 \rangle$, where $\langle \cdot \rangle$ denotes a time average over all valid time points $t$. To reduce bias arising from the finite trajectory length and ensure statistical robustness, the lag time $\tau$ was restricted to 30% of the total acquisition duration. The anomalous exponent $\alpha$ and the generalized diffusion coefficient $D_\alpha$ [µm² s⁻ᵅ] were obtained by fitting the 3D anomalous diffusion model $MSD(\tau) = 6D_\alpha \tau^\alpha$ in log-log space, i.e., by linear regression of log(MSD) versus log($\tau$) over the fitting range of 10-80 % of the computed lag-time. Parameter uncertainty was quantified using 95% confidence intervals derived from case bootstrap resampling of the (log($\tau$), log(MSD)) pairs within the fitting range. For each trajectory, 500 bootstrap resamples were generated, and the regression was refitted for each resample. Confidence intervals for the anomalous exponent $\alpha$ and generalized diffusion coefficient $D_\alpha$ were obtained from the corresponding percentile distributions. In Fig. 4e, the anomalous diffusion exponent estimated for each trajectory was spatially mapped to the particle's initial position, defined as its location in the first acquired frame.

**Data, Materials, and Software Availability**
The data provided in the manuscript is available from the corresponding author upon reasonable request.

**Acknowledgments**
This work was financially supported by Japan Society for the Promotion of Science (23H00273, 25H01386, T.I.), JST FOREST Program (JPMJFR236C, T.I.), Precise Measurement Technology Promotion Foundation (T.I.), UTEC-UTokyo FSI Research Grant (T.I.), RIKEN TRIP initiative (T.I.). M.F. was supported by the Japan Society for the Promotion of Science Research Fellowship for Young Scientists (DC1, 25KJ1144), the University of Tokyo MERIT-WINGS Program, and the JSR Fellowship.

**Author contributions**
M.F., Y.S. and S. K. developed the microscopy system. M.F. performed the experiments and analyzed the data. M.F., K.T., and T. I. discussed the interpretation of the results. K.T. and T.I. supervised the work. M.F., K.T., and T.I. wrote the manuscript with inputs from other authors.

**Competing interests**
K.T. and T.I. are inventors of patents related to the MIP-ODT system.